\documentclass[twocolumn, aps, prd, showpacs, nofootinbib, superscriptaddress]{revtex4}

\usepackage{amsmath}
\usepackage{graphicx}
\usepackage{hhline}
\usepackage[colorlinks=true, linkcolor=blue, anchorcolor=blue, citecolor=blue, urlcolor=blue]{hyperref}
\usepackage{lipsum}
\usepackage{multirow}
\usepackage{newtxmath}
\usepackage{pdflscape}
\usepackage{verbatim}
\usepackage{xcolor}
\definecolor{dark green}{rgb}{0.00, 0.39, 0.00}

\def\diff{\ensuremath{\mathrm{d}}}
\def\ff{\ensuremath{\mathtt{d_2\,}}}
\def\ffsq{\ensuremath{\mathtt{d_2}^2\,}}
\def\gg{\ensuremath{\mathtt{d_{3a}}}}
\def\ggg{\ensuremath{\mathtt{d_{3b}}}}

\newcommand{\tw}{\textwidth}
\newcommand{\cw}{\columnwidth}

\def\physrep{Phys.~Rep.}	% Physics Reports

\makeatletter
\def\url@rmstyle{%
  \@ifundefined{selectfont}{\def\UrlFont{\sf}}{\def\UrlFont{\footnotesize\rmfamily}}}
\makeatother
\def\aap{A\&A}			% Astronomy and Astrophysics
\def\aj{AJ}				% Astronomical Journal
\def\apj{ApJ}			% Astrophysical Journal
\def\apjl{ApJL}			% Astrophysical Journal, Letters
\def\apjs{ApJS}			% Astrophysical Journal, Supplement
\def\jcap{JCAP} 		%
\def\mnras{MNRAS}		% Monthly Notices of the RAS

\begin{document}

\title{Redshift-space distortion from dynamical dark energy \\with time-dependent Lagrangian perturbation theory}
%%% Authors
\author{Chakkrit Kaeonikhom}
\affiliation{Department of Physics and General Science, Chiang Mai Rajabhat University, Chiang Mai, 50300, Thailand}
\author{Teeraparb Chantavat}\thanks{E-mail: teeraparbc@nu.ac.th}
\affiliation{ThEP's Laboratory of Cosmology and Gravity, The Institute for Fundamental Study \\ ``The Tah Poe Academia Institute", Naresuan University, Phitsanulok, 65000, Thailand}
\affiliation{Thailand Center of Excellence in Physics, Ministry of Education, Bangkok 10400, Thailand}

\date{\today}

\begin{abstract}
We apply the Lagrangian perturbation theory with time-dependent growth functions at second and third order of perturbation with the aim to model the effect of dynamical dark energy on redshift-space distortions.  Our fiducial galaxy redshift surveys are modeled after the upcoming SKA and DESI redshift surveys.  We include PLANCK CMB priors and the 20\% uncertainty on the linear bias parameter, incorporating the unknown instrumentation noise.  After the marginalizing, our results show that the constraints on the density parameter could potentially get better than $\sim$1\%, while the constraints on $w_0$ and $w_{\rm a}$ could be better than $\sim 5\%$ and $\sim 60\%$ respectively, consistent with works done before.  However, the inclusion of time-dependent growth functions would alter the nonlinear power spectrum by as much as $\sim$10\%.   The inclusion of the time-dependent growth functions become crucial as the precision gets better.

\end{abstract}

\maketitle
\section{Introduction}
\label{sec:introduction}

Current cosmological data, such as the observed distance modulus of the type Ia supernova (SNIa) \citep{Riess_ea1998, Perlmutter_ea1999, Efstathiou_ea2002}, the cosmic microwave background (CMB) \citep{Jaffe_ea2001, Pryke_ea2002, Spergel_ea2007, Planck_2015}, the baryonic acoustic oscillations (BAOs) \citep{Eisenstein_ea2005} and an independent Hubble parameter measurement \citep{Freedman_ea2012} suggest that the expansion of the Universe is accelerating.  One of the explanations of the cause of the acceleration is the mysterious form of energy called ``dark energy."  The simplest model of dark energy is the cosmological constant \citep{Carroll_Press1992}.  While the cosmological constant is enough to explain the current data, its constancy leads to a fine-tuning problem \citep{Peebles_Ratra2003, Padmanabhan_2003}.  Apart from the cosmological constant, an alternative model of dark energy is an ideal fluid with an equation of state given by $P = w\rho c^2$, where $P$ and $\rho$ are fluid pressure and density, respectively.  $w$ is the equation of state which could, in general, vary with time.  In order for the fluid to give rise to the accelerating expansion, the value of $w$ must be $w \leqslant -\frac{1}{3}$.  The energy consideration requires $w \geqslant -1$; however, the current constraint on $w$ is very close to $-1$ \citep{Planck_2015}.  Therefore, distinguishing the cosmological constant and dynamical dark energy is a challenging problem in cosmology.  With the redshift surveys, the number of observed galaxies of more than $1\times10^6$ galaxies can potentially be able to constrain cosmological parameters especially the equation of state for the dynamical dark energy.  Since we can only observe in redshift space along the line of sight, the redshift-space distortion (RSD) effect is an inevitable consequence.  The RSD effect is sensitive to the density growth functions, which sequentially depend on $w$.  Hence,  galaxy surveys are viable probes of dark energy and accurate theoretical modeling of the RSD effect is crucial.

One of the prominent probes of the non linear structure of the Universe is baryonic acoustic oscillations, primordial sound waves which were propagated in the hot plasma of photons and baryons in the early Universe \citep{Eisenstein_Wu1998}.  The BAO effect leaves a distinct statistical imprint of oscillatory features in the power spectrum on the scale that corresponds to the sound horizon scale.  The BAO effect could be detected through the observed two-point correlation functions and the power spectra in redshift space \citep{Eisenstein_ea2005, Cole_ea2005, Tegmark_ea2006, Percival_ea2007, Percival_ea2010, Kazin_ea2010, Beutler_ea2011}.  The two-point correlation function provides a powerful probe of the time-dependent equation of state of dark energy model in a manner that is highly complementary to measurements of the cosmic microwave background.

Within the present time and the next few decades, there will be several large-scale-structure surveys that explore increasingly larger and deeper regions of the Universe; for example, the Dark Energy Survey (DES)\footnote{\url{http://www.darkenergysurvey.org}} \citep{DES}, the Extended Baryon Oscillation Spectroscopic Survey (eBOSS)\footnote{\url{http://www.sdss.org/surveys/eboss/}}, Euclid\footnote{\url{http://sci.esa.int/euclid/}}, the Large Synoptic Survey Telescope (LSST)\footnote{\url{http://www.lsst.org/}} \citep{LSST2009}, the Square Kilometre Array (SKA)\footnote{\url{https://www.skatelescope.org}}, and the Wide-Field InfraRed Survey Telescope (WFIRST)\footnote{\url{http://wfirst.gsfc.nasa.gov/}} \citep{Green_ea2012}.  These surveys will help our understanding of the growth of the structure in the Universe, especially in the nonlinear regimes.  Therefore, it is essential to have an accurate theoretical modeling to interpret the upcoming observational data. The futuristic surveys are crucial for observations of the BAOs because those surveys will map out to $z \gtrsim 3$.  At high redshifts, many more oscillatory modes will be available for observations and data analysis.  However, even at low redshift, for large-sky coverage surveys, it is marginally possible to observe the BAOs.

Analytical modeling of the large-scale structure of the Universe on non linear scales has the potential to considerably increase the science return of upcoming surveys by increasing the number of modes available for model comparisons. One way to achieve this is to model nonlinear scales perturbatively.   Therefore, predicting the precise nonlinear behavior of the galaxy power spectrum using analytical approaches and comparison to $N$-body simulations is an essential step in interpreting these data and in elucidating the nature of dark energy.  There are two main perturbative modelings in the literature, the Eulerian standard perturbation theory (SPT) \citep{Goroff_ea1986, Jain_Bertschinger1994, Scoccimarro_Frieman1996} and the Lagrangian perturbation theory (LPT) \citep{Zeldovich1970, Bouchet_Colombi1995, Matsubara2008}; for a review see Ref.~\cite{Bernardeau_ea2002}.

SPT requires two fundamental random fields; the overdensity field and the velocity field.  However, the convergence rate of SPT is slow, which comes at a great computational cost for higher accuracy by adding more terms.  In addition,  the higher-order SPT does not give any sensible prediction for the correlation function because unphysical behavior in the small wavelength limit prevents the Fourier transform from converging \citep{Carlson_ea2009}.  Therefore, in this article, we shall exploit the LPT resummation as an alternative to SPT.  LPT has an advantage over SPT because it is convenient and customary to set up the initial conditions for $N$-body simulations.  The calculation of the effects of redshift-space distortion is straightforward in LPT \citep{Taylor_Hamilton1996}.  In addition, LPT focuses on perturbing the displacement field rather than the overdensity and velocity fields.

The goal of this article is to investigate the potential of utilizing galaxy redshift surveys to constraint the equation of state parameter for dark energy in the BAO power spectrum in redshift space by applying an analytic cosmological perturbation theory (LPT) formalism in Ref.~\cite{Matsubara2008}.  In Sec.~\ref{sec:theory}, wed describe the LPT theory in both real space and redshift space with time dependent growth functions.  In Sec.~\ref{sec:models}, we describe the models and assumptions for our analysis and give the result in Sec.~\ref{sec:results}.  Discussions and conclusions are given in Sec.~\ref{sec:conclusion}.

\section{Theory}
\label{sec:theory}

In this section, we describe all the relevant theories in this analysis such as the structure formation theory \citep{Peebles1980}, Lagrangian perturbation theory, and the redshift-space distortion.  For a complete and more rigorous review of cosmological perturbation theory, we advise the reader to study Ref.~\cite{Bernardeau_ea2002}.  In this work, we follow the Lagrangian perturbation formalism in Ref.~\cite{Matsubara2008}.

\subsection{Structure formation theory}
\label{ssec:struc}

The structure that we see in the Universe is generated from small inhomogeneities where the fluctuation of the density field $\rho(\boldsymbol{x}, t)$ is defined as
\begin{equation}
	\delta(\boldsymbol{x}, t) \equiv \frac{\rho(\boldsymbol{x}, t) - \bar{\rho}(t)}{\bar{\rho}(t)},
\end{equation}
where $\delta(\boldsymbol{x}, t)$ is the \textit{density overdensity} or \textit{contrast} and $\boldsymbol{x}$ is the comoving distance vector.  $\bar{\rho}(t)$ is the average density in the comoving coordinate system.  The dynamics of the density field is governed by the continuity equation,
\begin{equation}
	\frac{\partial}{\partial t}\rho(\boldsymbol{x}, t) + \boldsymbol{\nabla}\cdot \bigg(\rho(\boldsymbol{x}, t)\, \boldsymbol{u}(\boldsymbol{x}, t) \bigg) = 0,
\end{equation}
and Euler's equation,
\begin{equation}
	\frac{\partial}{\partial t}\boldsymbol{u}(\boldsymbol{x}, t) + 2 H(t) \boldsymbol{u} (\boldsymbol{x}, t) + \boldsymbol{u} (\boldsymbol{x}, t) \cdot \boldsymbol{\nabla}\boldsymbol{u} (\boldsymbol{x}, t) = -\boldsymbol{\nabla}\Phi (\boldsymbol{x}, t),
\end{equation}
where $\boldsymbol{u}$ is the \textit{peculiar velocity} and $\boldsymbol{\nabla} \equiv \partial_{\boldsymbol{x}}$ is the comoving gradient operator.  $H(t) \equiv \dot{a}/a$ is the Hubble expansion rate.  $\Phi$ is the comoving gravitational potential,
\begin{equation}
	\label{eq:poisson}
	\nabla^2 \Phi(\boldsymbol{x}, t) = \frac{3}{2} \Omega_{\rm M}(t) H^2(t) \delta(\boldsymbol{x}, t),
\end{equation}
where $\Omega_{\rm M}$ is the \textit{matter density parameter}.

We shall define the \textit{velocity potential}, $\theta$, as the divergence of the comoving peculiar velocity field
\begin{equation}
	\theta(\boldsymbol{x}, t) \equiv \boldsymbol{\nabla}\cdot\boldsymbol{u}.
\end{equation}
In the linear cosmological perturbation theory, the growth of the density and the velocity potential are given by
\begin{equation}
	\label{eq:continuity}
	\frac{\partial}{\partial t}\delta(\boldsymbol{x}, t) + \theta(\boldsymbol{x}, t)= 0,
\end{equation}
and
\begin{equation}
	\label{eq:euler}
	\frac{\partial}{\partial t} \theta(\boldsymbol{x}, t) + 2 H\, \theta(\boldsymbol{x}, t) = \frac{3}{2}\Omega_{\rm M}(t) H^2(t) \delta(\boldsymbol{x}, t).
\end{equation}
From Eqs.~(\ref{eq:continuity}) and (\ref{eq:euler}), a second-order differential equation in terms of the overdensity is given by,
\begin{equation}
	\label{eq:diffdelta}
	\frac{\partial^2}{\partial t^2}\delta(\boldsymbol{x}, t) + 2 H(t) \frac{\partial}{\partial t} \delta(\boldsymbol{x}, t) =  \frac{3}{2}\Omega_{\rm M}(t) H^2(t) \delta(\boldsymbol{x}, t).
\end{equation}
Assuming a linear growth factor $D_1(t)$ such that
\begin{equation}
	\label{eq:defgrowth}
	\delta(\boldsymbol{x}, t) = D_1(t) \delta_0(\boldsymbol{x}),
\end{equation}
where $\delta_0(\boldsymbol{x})$ is the overdensity at a particular time.  It is customary to normalize the growth function to unity so that $\delta_0(\boldsymbol{x})$ represent the overdensity at the present epoch.  From Eqs.~(\ref{eq:diffdelta})--(\ref{eq:defgrowth}), we have the equations describing the evolution of the linear density perturbations as
\begin{equation}
	\ddot{D}_1 + 2 H(t) \dot{D}_1 - \frac{3}{2} \Omega_{\rm M}(t) H^2(t) D_1(t) = 0.
\end{equation}
$D_1$ is known as the linear growth factor and has been used pretty much in a linear regime and as an extrapolation to nonlinear regimes.  Perturbatively, we can expand the overdensity as
\begin{equation}
	\label{eq:perturb1}
	\delta(\boldsymbol{x}, t) \equiv \delta_1(\boldsymbol{x}, t) + \delta_2(\boldsymbol{x}, t) + \delta_3(\boldsymbol{x}, t) + \ldots,
\end{equation}
where $\delta_n(\boldsymbol{x}, t)$ is proportional to $\delta_0^n(\boldsymbol{x})$.  Similarly, we can define
\begin{equation}
	\label{eq:perturb2}
	\Phi(\boldsymbol{x}, t) \equiv \Phi_1(\boldsymbol{x}, t) + \Phi_2(\boldsymbol{x}, t) + \Phi_3(\boldsymbol{x}, t) + \ldots,
\end{equation}
where, from Eq.~(\ref{eq:poisson}), $\nabla^2\Phi_n(\boldsymbol{x}, t) = (3/2) \Omega_{\rm M} H^2 \delta_n(\boldsymbol{x}, t)$.  In addition,
\begin{equation}
	\label{eq:perturb3}
	\theta(\boldsymbol{x}, t) \equiv \theta_1(\boldsymbol{x}, t) + \theta_2(\boldsymbol{x}, t) + \theta_3(\boldsymbol{x}, t) + \ldots,
\end{equation}
where $\theta_n(\boldsymbol{x}, t)$ is proportional to $\delta_0^n(\boldsymbol{x})$.  With perturbative definitions in Eqs.~(\ref{eq:perturb1})--(\ref{eq:perturb3}), additional growth functions in higher order are given by
\begin{eqnarray}
	\label{eq:growth2}
	\ddot{D}_2 + 2 H \dot{D}_2 - \frac{3}{2}\Omega_{\rm M} H^2 D_2 &=& \frac{3}{2}\Omega_{\rm M} H^2 D_1^2, \\
	\label{eq:growth3-1}
	\nonumber\ddot{D}_{3a} + 2 H \dot{D}_{3a} - \frac{3}{2}\Omega_{\rm M} H^2 D_{3a} &=& \frac{3}{2}\Omega_{\rm M} H^2 \left(D_1^3 + D_1D_2 \right), \\
	&& \\
	\label{eq:growth3-2}
	\ddot{D}_{3b} + 2 H \dot{D}_{3b} - \frac{3}{2}\Omega_{\rm M} H^2 D_{3b} &=& \frac{3}{2}\Omega_{\rm M} H^2 D_1^3. 
\end{eqnarray}
We define $D_2(t)$ as the second-order growth function, and $D_{3a}(t)$ and $D_{3b}(t)$ as the third-order growth functions --- there are two independent third-order growth functions.  Notice that $D_2(t) \propto D_1^2(t)$ and $D_{3a}(t), D_{3b}(t) \propto D_1^3(t)$; hence, it is convenient to define $\displaystyle \ff(t) \equiv D_2(t) / D_1^2(t)$, $\displaystyle \gg(t) \equiv D_{3a}(t) / D_1^3(t)$ and $\displaystyle \ggg(t) \equiv D_{3b}(t) / D_1^3(t)$.  For an Einstein--de Sitter (EdS) universe, $\ff = 3/7$, $\gg = 5/21$ and $\ggg = 1/6$.

\subsection{Lagrangian perturbation theory}
\label{ssec:lpt}

We shall follow the Lagrangian perturbation theory from Ref.~\cite{Matsubara2008}.  In this model of perturbation, the Eulerian particle position $\boldsymbol{x}$ can be mapped to a Lagrangian particle position $\boldsymbol{q}$ by the displacement field $\boldsymbol{\Psi}$,
\begin{equation}
	\boldsymbol{x}(t) = \boldsymbol{q}(\boldsymbol{x}) + \boldsymbol{\Psi}(\boldsymbol{q}, t),
\end{equation}
where $\boldsymbol{x}$ is in the Eulerian comoving coordinate description.  The overdensity in the Lagrangian position is given by
\begin{equation}
	\label{eq:overdensity}
	\delta(\boldsymbol{x}, t) = \int \diff^3 q\ \delta_{\rm D}^3(\boldsymbol{x} - \boldsymbol{q} - \boldsymbol{\Psi}(\boldsymbol{q})) - 1,
\end{equation}
where $\delta_{\rm D}^3(\boldsymbol{x})$ is the Dirac's delta function in three dimensions.  The equation of motion in terms of the displacement field is given by
\begin{equation}
	\frac{\diff^2}{\diff t^2} \boldsymbol{\Psi} (\boldsymbol{q}, t) + 2 H(t) \frac{\diff}{\diff t} \boldsymbol{\Psi} (\boldsymbol{q}, t) = -\boldsymbol{\nabla} \Phi(\boldsymbol{x}, t).
\end{equation}
In principle, we can expand $\boldsymbol{\Psi}(\boldsymbol{q}, t)$ as
\begin{equation}
	\boldsymbol{\Psi}(\boldsymbol{q}, t) = \boldsymbol{\Psi}_1(\boldsymbol{q}, t) + \boldsymbol{\Psi}_2(\boldsymbol{q}, t) + \boldsymbol{\Psi}_3(\boldsymbol{q}, t) + \ldots.
\end{equation}
We can define the Fourier mode of the displacement field $\boldsymbol{\Psi}$ as
\begin{eqnarray}
	\nonumber\boldsymbol{\widetilde{\Psi}}_n(\boldsymbol{k}, t) &=& \frac{\mathrm{i}D_1^n(t)}{n!} \left( \prod_{i = 1}^n \int \frac{\diff^3 k_i}{(2\pi)^3}\right) \left(2\pi \right)^3 \delta_{\rm D}^3\left( \sum_{i=1}^n \boldsymbol{k}_i - \boldsymbol{k} \right) \\
	&& \times\ \boldsymbol{L}_n(\boldsymbol{k}_1, \ldots, \boldsymbol{k}_n, t) \left( \prod_{i = 1}^n\tilde{\delta}_0(\boldsymbol{k}_i)\right),
\end{eqnarray}
where $\boldsymbol{\widetilde{\Psi}}_n(\boldsymbol{k}, t)$ is the $n$th order displacement vector field in Fourier space and $\boldsymbol{L}_n(\boldsymbol{k}_1, \ldots, \boldsymbol{k}, t)$ are the Lagrangian perturbative kernels.  Our convention for the Fourier transform is
\begin{equation}
	\tilde{f}(\boldsymbol{k}) = \int\diff^3 x\ f(\boldsymbol{x}) \exp\bigg( -\mathrm{i} \boldsymbol{k} \cdot \boldsymbol{x}\bigg),
\end{equation}
and
\begin{equation}
	f(\boldsymbol{x}) = \int\frac{\diff^3 k}{(2\pi)^3}\ \tilde{f}(\boldsymbol{k}) \exp \bigg(\mathrm{i} \boldsymbol{k} \cdot \boldsymbol{x} \bigg).
\end{equation}
The time-dependent Lagrangian perturbative kernels are given by
\begin{widetext}
\begin{eqnarray}
	\boldsymbol{L}_1(\boldsymbol{k}_1, t) &=& \frac{\boldsymbol{k}}{k^2}, \\
	\boldsymbol{L}_2(\boldsymbol{k}_1, \boldsymbol{k}_2, t) &=& \ff(t) \frac{\boldsymbol{k}}{k^2} \left[ 1 - \frac{\left( \boldsymbol{k}_1 \cdot \boldsymbol{k}_2 \right)^2}{k_1^2 k_2^2} \right], \\
	\nonumber\boldsymbol{L}_{3a}(\boldsymbol{k}_1, \boldsymbol{k}_2, \boldsymbol{k}_3, t) &=& 3 \gg(t) \left(\frac{\boldsymbol{k}}{k^2}\right) \left( 1 - 3 \frac{\left(\boldsymbol{k}_1 \cdot \boldsymbol{k}_2 \right)^2}{k_1^2 k_2^2} \right) \left( 1 - \frac{\left( \left( \boldsymbol{k}_1 + \boldsymbol{k}_2 \right) \cdot \boldsymbol{k}_3 \right)^2}{1} \right) \\
	&& -2 \ggg(t) \left(\frac{\boldsymbol{k}}{k^2}\right) \left[ 1 - 3 \frac{\left( \boldsymbol{k}_1 \cdot \boldsymbol{k}_2 \right)^2}{k_1^2 k_2^2} + 2 \frac{\left( \boldsymbol{k}_1 \cdot \boldsymbol{k}_2 \right) \left( \boldsymbol{k}_2 \cdot \boldsymbol{k}_3 \right) \left( \boldsymbol{k}_3 \cdot \boldsymbol{k}_1 \right)}{k_1^2 k_2^2 k_3^2} \right],
\end{eqnarray}
\end{widetext}
where $\boldsymbol{k} = \displaystyle\sum_i \boldsymbol{k}_i$.  $L_{3a}$ will need to be symmetrized in terms of $\boldsymbol{k}$'s later.

\begin{figure}
	\includegraphics[width=1.0\cw]{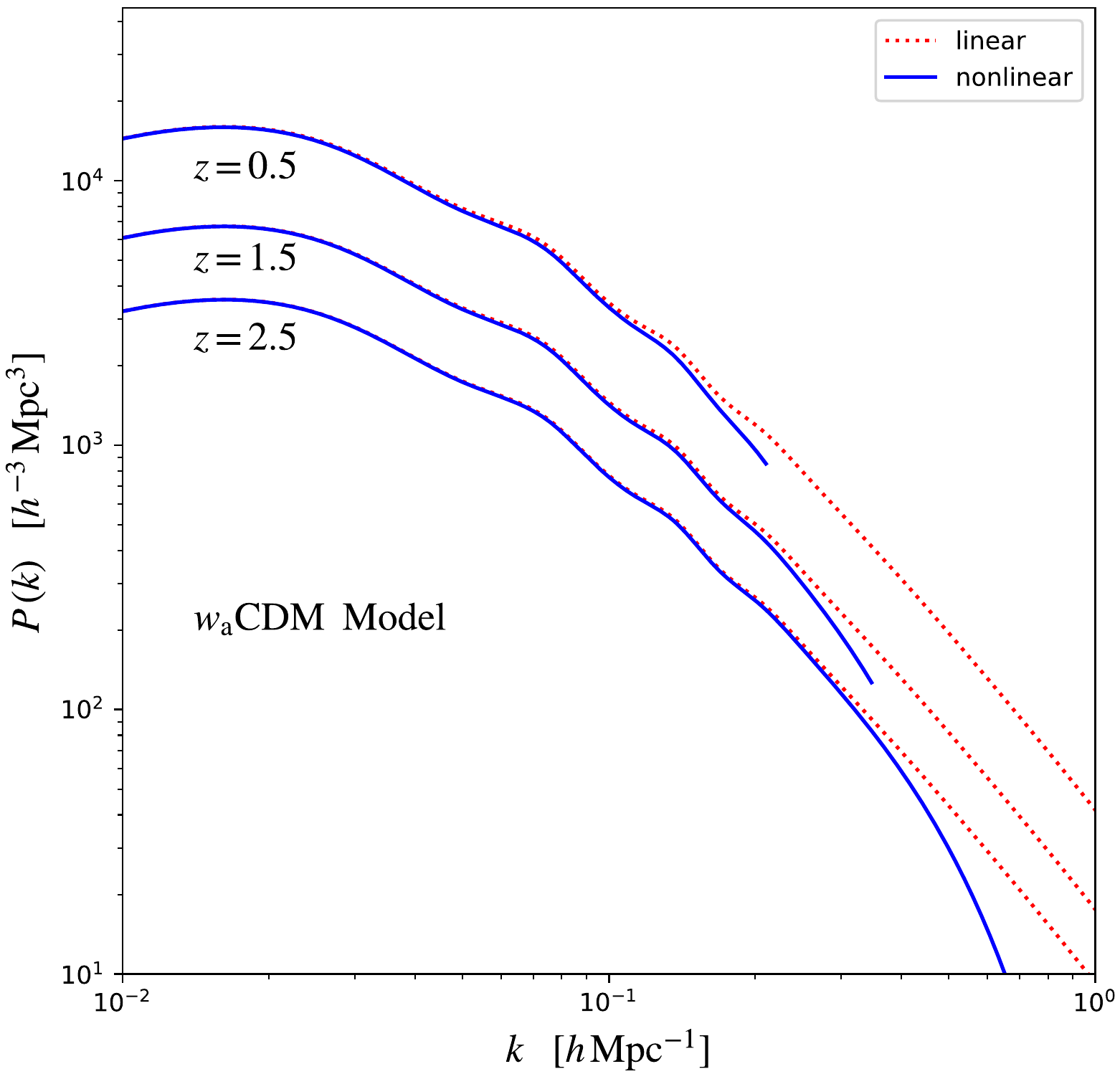}
	\caption{\label{fig:powerspectra} The linear matter power spectra (dotted) and nonlinear matter power spectra (solid) at different redshifts in the $w_{\rm a}$CDM model (see Table~\ref{tab:modelsparams}).  The same plots for the other models have similar features.  From top to bottom, $z = 0.5, 1.5$, and $2.5$.  The nonlinear power spectra are shown up to $k_{\rm nl}$; however, we apply a cutoff scale at $k_{\rm nl}/2$.
}
\end{figure}

\subsubsection{Real-space power spectrum}
\label{sssec:real}

The power spectrum is defined as a Fourier transform of the two-point correlation function of overdensity,
\begin{equation}
	P(k) \equiv \int\diff^3 x\ \xi(x) \exp(-\mathrm{i} \boldsymbol{k} \cdot \boldsymbol{x}),
\end{equation}
where $\xi$ is the two-point correlation function.  The power spectrum has a simple relation as
\begin{equation}
	\bigg\langle\tilde{\delta}(\boldsymbol{k}) \tilde{\delta}^*(\boldsymbol{k}^\prime) \bigg\rangle = (2\pi)^3\delta_{\rm D}^3(\boldsymbol{k} - \boldsymbol{k}') P(k),
\end{equation}
where $P(k)$ is the power spectrum and $\tilde{\delta}(\boldsymbol{k})$ is the Fourier transform of overdensity [Eq.~(\ref{eq:overdensity})].  The power spectrum in Fourier space depends only on the magnitude of the wave vector, $k = |\boldsymbol{k}|$.  Following the formalism in Ref.~\cite{Matsubara2008}, the power spectrum is given by
\begin{align}
	\label{eq:realps}
	\nonumber P(k, z) &= \exp\left[-\frac{k^2}{6\pi^2}\int\!\mbox{d}p\ P_{\rm L}(p, z)\right] \\
	\nonumber&\quad \times \Bigg\{ P_{\rm L}(k, z) + \left( \frac{\ffsq\!(z)}{2}Q_1(k) + \ff\!(z) Q_2(k) + \frac{1}{2} Q_3(k) \right) \\
	&\quad+ 2 \bigg( \gg(z) R_1(k) + \ff\!(z) R_2(k)\bigg) \Bigg\},
\end{align}
where $P_L(k)$ is the linear matter power spectrum. $Q_n(k)$'s and $R_n(k)$'s are the integral functionals of the form
\begin{eqnarray}
	\label{eq:qn}
	\nonumber Q_n(k) &=& \frac{k^3}{4\pi^2} \int_0^\infty \diff r\ P_{\rm L}(kr) \int_{-1}^{1} \diff x\\
	&& \times\ P_{\rm L}\left[ k(1 + r^2 - 2rx)^{1/2}\right] \frac{\tilde{Q}_n(r, x)}{(1 + r^2 - 2rx)^2}, \\
	\label{eq:rn}
	R_n(k) &=& \frac{1}{48} P_{\rm L}(k) \frac{k^3}{4\pi^2} \int_0^\infty \diff r\ P_{\rm L}(kr)\ \tilde{R}_n(r),
\end{eqnarray}
where $P_L(k) \equiv P_L(k, z=0)$ the linear power spectrum at the present epoch.
\begin{eqnarray}
	\tilde{Q}_1(r,x) &=& r^2\left(1 - x^2\right)^2, \\
	\tilde{Q}_2(r,x) &=& \left(1-x^2\right)rx\left(1-rx\right), \\
	\tilde{Q}_3(r,x) &=& x^2\left(1-rx\right)^2, \\
	\tilde{Q}_4(r,x) &=& 1 - x^2,
\end{eqnarray}
and
\begin{eqnarray}
	\nonumber\tilde{R}_1(r) &=& -\frac{2}{r^2}\left(1 + r^2\right)\left(3 - 14r^2 + 3r^4\right) \\
	&& +\ \frac{3}{r^3}\left(r^2 - 1\right)^4 \ln\left|\frac{1 + r}{1 - r}\right|, \\
	\label{eq:r2}
	\nonumber \tilde{R}_2(r) &=& \frac{2}{r^2}\left(1 - r^2\right)\left(3 - 2r^2 + 3r^4\right) \\
	&& +\ \frac{3}{r^3}\left(r^2 - 1\right)^3\left(1 + r^2 \right) \ln\left|\frac{1 + r}{1 - r}\right|.
\end{eqnarray}
We shall express time in terms of redshift $z$ for observables.  For an Einstein--de Sitter universe, our result is consistent with Ref.~\cite{Matsubara2008}.  The validity of this approach is applicable when the argument of the exponential term in Eq.~(\ref{eq:realps}) is of the order of unity; hence, we define the nonlinear scale $k_{\rm nl}$ as
\begin{equation}
	k_{\rm nl} \equiv \left[ \frac{1}{6\pi^2}\int \diff k' P_{\rm L}(k')\right]^{-1/2}.
\end{equation}
We will apply a cutoff scaling where $k < k_{\rm nl}/2$.  The power spectra for the $w_{\rm a}$CDM model (see Sec.~\ref{sec:models}) is shown in Fig.~\ref{fig:powerspectra} as an example.

\subsubsection{Redshift-space power spectrum}
\label{sssec:redshift}

The power spectrum in redshift space is anisotropic and, hence, depends on the observed direction $\hat{z}$, the radial direction.  Following the formalism in Ref.~\cite{Matsubara2008}, the redshift-space power spectrum is given by
\begin{eqnarray}
	\nonumber P_{\rm s}(\boldsymbol{k}, \mu, z) &=& \exp\left\{ -k^2 \left[1 + f(f + 2)\mu^2 \right] A \right\} \\
	\nonumber && \times \left[ (1 + f \mu^2)^2 P_{\rm L}(k, z) + \sum_{n,m}\mu^{2n} f^m E_{nm}(k) \right],\\
	\label{eq:ps}
\end{eqnarray}
where $\mu = \hat{z} \cdot \boldsymbol{k}/k$ is the direction cosine of the wave vector $\boldsymbol{k}$ with respect to the line of sight,
\begin{equation}
	A = \frac{1}{6\pi^2} \int \mbox{d}k\ P_L(k),
\end{equation}
and
\begin{equation}
	f = \frac{\mbox{d} \ln D_1}{\mbox{d}\ln a}.
\end{equation}
The integral $E_{mn}(k)$'s are the integral functionals in terms of $Q_n(k)$ and $R_n(k)$ of the forms
\begin{eqnarray}
	\nonumber E_{00}(k) &=& \frac{\ffsq\!(t)}{2} Q_1(k) + \ff\!(t) Q_2(k) + \frac{1}{2} Q_3(k) \\
	&& + 2 \gg(t) R_1(k) + 2 \ff\!(t) R_2(k), \\
	E_{11}(k) &=& 4 E_{00}(k), \\
	\nonumber E_{12}(k) &=& -\frac{\ff\!(t)}{2} Q_1(k) -\frac{3}{2}Q_2(k) + \frac{1}{4} Q_4(k) \\
	&& - 2 \ff\!(t) R_1(k), \\
	\nonumber E_{22}(k) &=& \left(2\ffsq\!(t) + \frac{\ff\!(t)}{2} \right) Q_1(k) + \left(5\ff\!(t) + \frac{3}{2} \right) Q_2(k) \\
	\nonumber && + 3 Q_3(k) -\frac{1}{4}Q_4(k)  + \bigg(2 \ff\!(t) + 6 \gg(t) \bigg) R_1(k) \\
	&& + 10 \ff\!(t) R_2(k),\\
	\nonumber E_{23}(k) &=& -\ff\!(t) Q_1(k) - 3 Q_2(k) + \frac{1}{2} Q_4(k) \\
	&& - 2 \ff\!(t) R_1(k),\\
	E_{24}(k) &=& \frac{3}{16}Q_1(k), \\
	\nonumber E_{33}(k) &=& \ff\!(t) Q_1(k) + \bigg(2 \ff\!(t) + 3 \bigg) Q_2(k) + 2 Q_3(k) \\
	&&- \frac{1}{2} Q_4(k) + 2 \ff\!(t) R_1(k) + 4 \ff\!(t) R_2(k), \\
	E_{34}(k) &=& - \frac{3}{8}Q_1(k) - \frac{3}{2}Q_2(k) + \frac{1}{4}Q_4(k), \\
	E_{44}(k) &=& \frac{3}{16}Q_1(k) + \frac{3}{2}Q_2(k) + \frac{1}{2}Q_3(k) - \frac{1}{4}Q_4(k).
\end{eqnarray}
For an EdS universe, $\ff = 3/7$, $\gg = 5/21$, and $\ggg = 1/6$, we recover the results done by Ref.~\citep{Matsubara2008}.

\section{Models and Data Analysis}
\label{sec:models}

In galaxy redshift surveys, the peculiar velocities of galaxies cause a distortion between the observed power spectrum (or correlation function) along a line of slight and the direction perpendicular to the line of slight.  The distortion effect can be predicted and tested with the underlying cosmology known as the Alcock-Paczy\'{n}ski (AP) test \citep{Alcock_Paczynski1979}.  The AP test can be used to track the effect of the dynamical dark energy in redshift surveys with different redshift slices.  In this section, we describe our dynamical dark energy models and analytical methods for galaxy redshift surveys.

\begin{table}
\caption{\label{tab:modelsparams} Fiducial cosmological parameters and their value for all models.}
\begin{tabular}{|p{0.22\cw}||p{0.21\cw}|p{0.21\cw}|p{0.21\cw}|}
	\cline{2-4}
	\multicolumn{1}{c||}{}& \hfil$\Lambda$CDM & \hfil $w$CDM & \hfil $w_{\rm a}$CDM \\
	\hline
	\hline
	\hfil $\Omega_{\rm B} h^2$ & \multicolumn{3}{|c|}{0.0223} \\
	\hline
	\hfil $\Omega_{\rm M} h^2$ &  \multicolumn{3}{|c|}{0.1401} \\
	\hline
	\hfil $\Omega_\Lambda$ &  \multicolumn{3}{|c|}{0.7019} \\
	\hline
	\hfil $\Delta_{\mathcal R}^2 \times 10^9$  &  \multicolumn{3}{|c|}{2.2061} \\
	\hline
	\hfil $\tau$&  \multicolumn{3}{|c|}{0.0956} \\
	\hline
	\hfil $n_s$&  \multicolumn{3}{|c|}{0.9695} \\
	\hline
	\hfil $w_0$ & \hfil -1.0 & \hfil -0.8 & \hfil -1.0 \\
	\hline
	\hfil $w_{\rm a}$ & \hfil 0.0 & \hfil 0.0 & \hfil 0.1 \\
	\hline
\end{tabular}
\end{table}

\subsection{Dynamical dark energy}

We use a parametrization of the equation of state of the dynamical dark energy by using a Taylor expansion in the Chevallier-Polarski-Linder (CPL) model \citep{Chevallier_Polarski2001, Linder_2003};
\begin{equation}
	w(a) = w_0 + w_{\rm a}(1 - a),
\end{equation}
or
\begin{equation}
	w(z) = w_0 + w_{\rm a} \frac{z}{1 + z},
\end{equation}
where $w_0$ and $w_{\rm a}$ are constants.  In the case of the cosmological constant, $w_0 = -1.0$ and $w_{\rm a} = 0.0$. In general the equation of state parameters $w_0$ and $w_{\rm a}$ could be any values allowable by observational data.  For the current observational constraint of $w_0$ and $w_{\rm a}$\footnote{At the time of preparation for the manuscript, the updated cosmological parameter constraints, PLANCK 2018, results have been released \citep{Planck_2018_06}.}, we use the MCMC exploring chain from {\tt planck\_{}lowl\_{}lowLike\_{}highL\_{}BAO} data with the base {\tt base\_{}w\_{}wa}.  This data set includes the constraints on parameters of our interest.  In addition, these data include late-time BAO observational data from DR11LOWZ, DR11CMASS \citep{SDSS_DR11}, MGS, and 6DF \citep{6ds_collaboration}.  From the data, we have
\begin{equation}
	\label{eq:planck_w0wa}
	w_0 = -1.030 \pm 0.361, \qquad w_{\rm a} = -0.334 \pm 0.909.
\end{equation}

\begin{table*}
\caption{\label{tab:surveys} The value of different survey parameters for each cosmological model (see text).}
\begin{tabular}{|p{0.1\tw}||p{0.1\tw}||p{0.1\tw}|p{0.1\tw}|p{0.1\tw}|p{0.1\tw}|}
	\cline{2-6}
	\multicolumn{1}{c||}{} & \hfil Survey & \hfil$z$ & \hfil $\bar{n}_g$ & \hfil $V_s$ & \hfil $b_1$ \\
	\multicolumn{1}{c||}{} & & & \hfil (Mpc$^{-3} h^3$) & \hfil (Mpc$^3 h^{-3}$) & \\
	\hline
	\hline
	& \hfil SKA1 & \hfil 0.0 -- 1.0 &\hfil$0.278$ & \hfil$3.80\times10^{10}$ & \hfil 1.052 \\
	\cline{2-6}
	\hfil $\Lambda$CDM & \hfil SKA2 & \hfil 1.0 -- 2.0 &\hfil$1.80\times10^{-3}$ & \hfil$1.09\times10^{11}$ & \hfil 2.091 \\
	\cline{2-6}
	&  \hfil DESI & \hfil 2.0 -- 3.0 & \hfil$2.06\times10^{-5}$ & \hfil$5.79\times10^{10}$ & \hfil 3.703 \\
	\hline
	\hline
	&  \hfil SKA1 & \hfil 0.0 -- 1.0 & \hfil$0.300$ & \hfil$3.34\times10^{10}$ & \hfil 1.049 \\
	\cline{2-6}
	\hfil $w$CDM &  \hfil SKA2 & \hfil 1.0 -- 2.0 & \hfil$2.07\times10^{-3}$ & \hfil$9.49\times10^{10}$ & \hfil 2.093 \\
	\cline{2-6}
	&  \hfil DESI & \hfil 2.0 -- 3.0 & \hfil$2.32\times10^{-5}$ & \hfil$5.79\times10^{10}$ & \hfil 3.701 \\
	\hline
	\hline
	&  \hfil SKA1 & \hfil 0.0 -- 1.0 & \hfil$0.279$ & \hfil$3.34\times10^{10}$ & \hfil 1.051 \\
	\cline{2-6}
	\hfil $w_{\rm a}$CDM &  \hfil SKA2 & \hfil 1.0 -- 2.0 & \hfil$1.83\times10^{-3}$ & \hfil$1.07\times10^{11}$ & \hfil 2.091 \\
	\cline{2-6}
	&  \hfil DESI & \hfil 2.0 -- 3.0 & \hfil$2.09\times10^{-5}$ & \hfil$5.71\times10^{10}$ & \hfil 3.703 \\
	\hline
\end{tabular}
\end{table*}

We demonstrate the application of the LPT formalism by considering the following fiducial models;  a $\Lambda$CDM-like model which follows the standard $\Lambda$CDM model with the cosmological constant ($w_0 = -1.0$ and $w_{\rm a} = 0.0$), $w$CDM-like models with $w_0 = -0.8$ and $w_{\rm a} = 0.0$, and a $w_{\rm a}$CDM-like model with $w_0 = -1$ and $w_{\rm a} = 0.1$.  In the $\Lambda$CDM model with cosmological constant $\Lambda$ both $w_0$ and $w_{\rm a}$ are constant; however, for academic proposes, we shall vary both parameters.  The summary of all the fiducial models and their parameters are summarized in Table~\ref{tab:modelsparams}.

\subsection{Galaxy bias}

The measurement of the power spectrum depends on the type of tracers; however, they may not exactly follow the distribution of the underlying dark matter.  For galaxy-type tracers on large scales, we shall apply a linear bias $b_1$ as \citep{Kaiser1984, Fry_Gaztanaga1993},
\begin{equation}
	\label{eq:bias}
	\delta_{g}(\boldsymbol{k}) \equiv b_1 \delta(\boldsymbol{k}),
\end{equation}
where $\delta_{g}(\boldsymbol{k})$ is the galaxy overdensity.  The power spectrum with galaxies as tracers is given by replacing $P_{\rm L}(k) \rightarrow b_1^2 P_{\rm L}(k)$ in Eq.~(\ref{eq:ps}),
\begin{equation}
	P_{\rm s, g}(\boldsymbol{k}, \mu, z) \equiv P_{\rm s}(\boldsymbol{k}, \mu, z)\bigg|_{P_{\rm L}(k) \rightarrow b_1^2 P_{\rm L}(k)}.
\end{equation}
The value of $b_1$ depends on surveys described in the next section.

\subsection{Galaxy surveys}

In this section, we give a brief introduction to some of the selected future surveys.  Additionally, we describe how those surveys will motivate us on how we decide on our fiducial surveys and redshift bins based on the real surveys.

\subsubsection{SKA}

The Square Kilometre Array (SKA) is a giant radio telescope located at two sites in South Africa and Western Australia.  The South African site will host midfrequency  receivers (350 MHz--15.3 GHz) while the Australia's site will host low-frequency receivers (50 MHz--350 MHz)\footnote{Data taken from SKA's Baseline Design document version2 (October 2015) at \url{https://www.skatelescope.org}}. The first operation phase will commence in 2020.  The second phase, which is about $10$ times more sensitive, will be active in about 2025.  One of the key science goals of SKA is to understand the nature of dark energy.  With the capabilities of SKA, a large number of HI galaxies could be detected with precise redshift information using the 21-cm line from the spin-flip transition.  Since the rest frequency of the transition is at 1420 MHz, telescopes with frequency ranges from $\sim$100 to 1420 MHz such as the SKA could detect HI galaxies up to $z \sim 10$ or the epoch of reionization.

For an SKA-like fiducial survey, we follow the bias and number count fitting functions for SKA from Ref.~\citep{Santos_ea2015} via a mock catalog \citep{Obreschkow_ea2009}.  The number count per redshift ${\rm d}n/{\rm d}z$ and the linear bias $b_1(z)$ as a function of redshift are given by
\begin{equation}
	{\rm d}n/{\rm d}z = 10^{6.7767}z^{2.1757}\exp\left(- 6.6374\, z\right),
\end{equation}
and
\begin{equation}
	b_1(z) = 0.5887 \exp\left( 0.813\,z\right),
\end{equation}
with parameters emulating the SKA phase 2 capabilities.

\begin{figure*}
	\includegraphics[width=\tw]{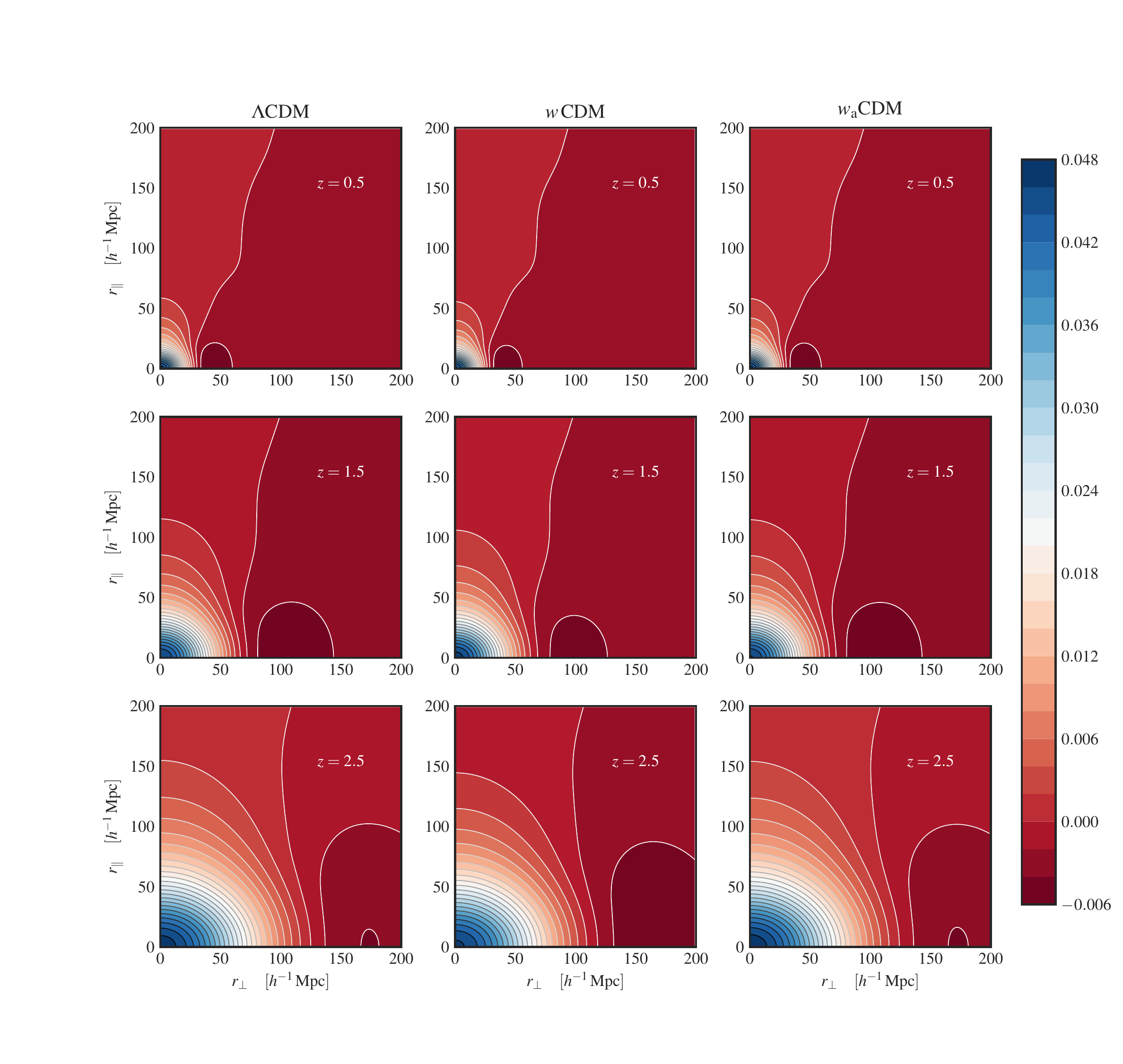}
	\caption{\label{fig:xi}The correlation function of the LPT nonlinear power spectrum in redshift space where $r_\perp$ is the perpendicular component to the line of sight and $r_\parallel$ is the parallel component to the line of sight.  The models are, from left to right, $\Lambda$CDM, $w$CDM, and $w_{\rm a}$CDM, respectively (see Table~\ref{tab:modelsparams}).  The bins are, from top to bottom, SKA1, SKA2, and DESI, respectively.}
\end{figure*}

\begin{figure}
	\includegraphics[width=\cw]{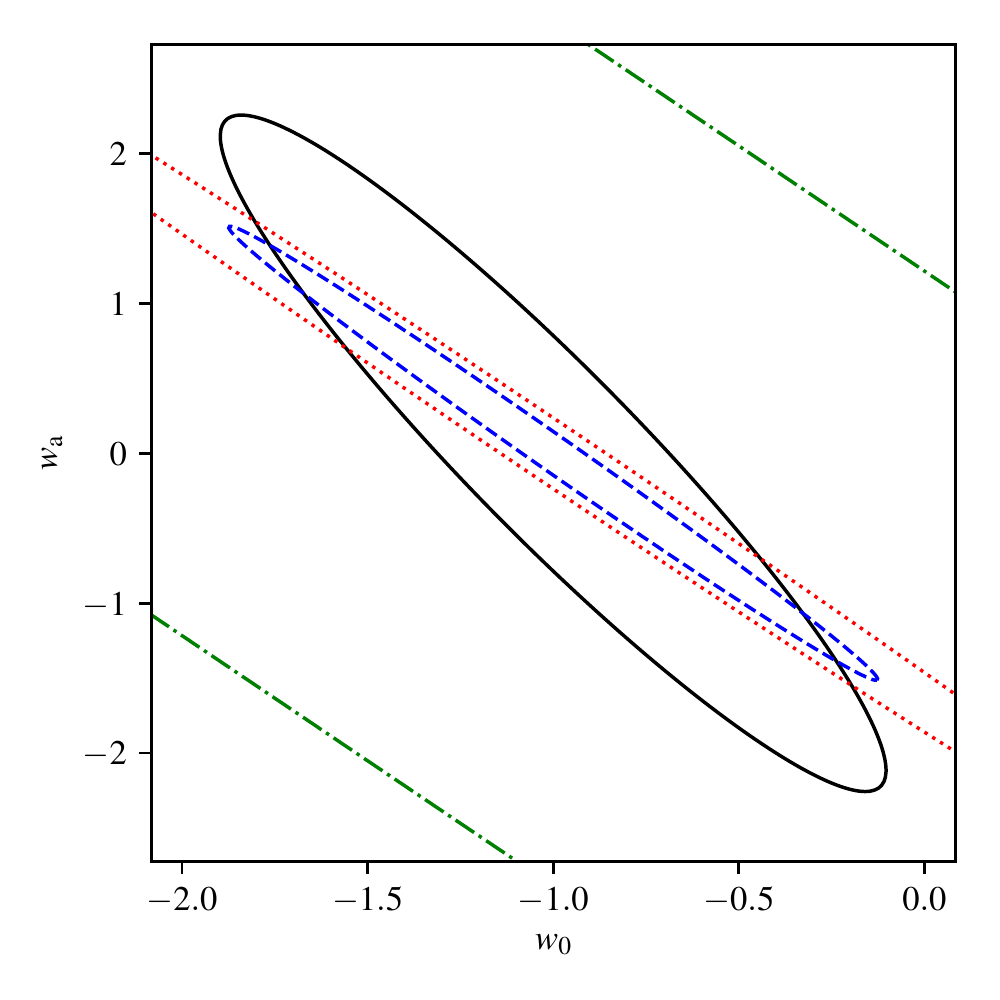}
	\caption{\label{fig:bins}An example of 95\% C.L. constraints on $w_0$ and $w_{\rm a}$ from different redshift bins for the $w_{\rm a}$CDM model.  The dark solid line is the constraint from PLANCK.  The innermost dashed blue line is the constraint from SKA1, the lowest redshift bin.  The dotted red line is the constraint from the SKA2 bin and the outermost dotted-dashed line is from the DESI bin.}
\end{figure}

\begin{figure*}
	\includegraphics[width=\tw]{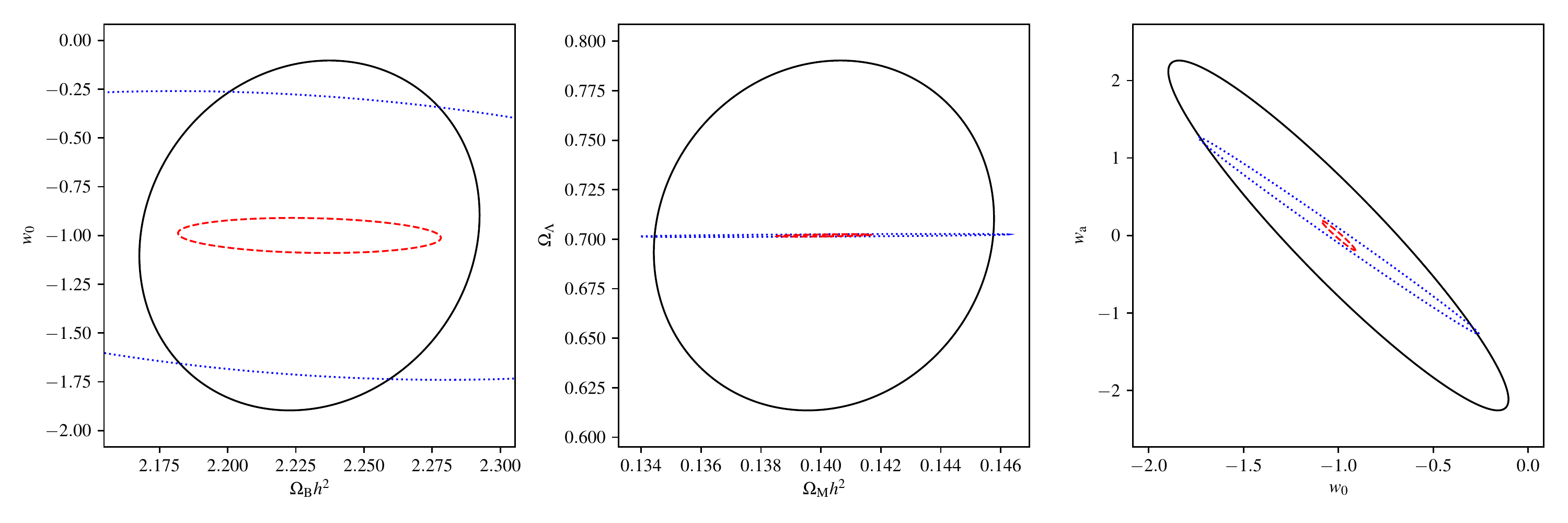}
	\caption{\label{fig:sigmaconstraint}95\% C.L. confidence level constraints on some of the cosmological parameter pairs; $\Omega_{\rm B} h^2$  and $w_0$ (left), $\Omega_{\rm M}$ and $\Omega_\Lambda$ (middle), $w_0$ and $w_{\rm a}$ (right) for the $w_{\rm a}$CDM model.  The dark solid line is the PLANCK constraints on the parameters from  {\tt planck\_{}lowl\_{}lowLike\_{}highL\_{}BAO} MCMC chain.  The blue dotted line is the constraints from galaxy surveys (all bins) and the red dashed line in the combined PLANCK and galaxy surveys constraints.}
\end{figure*}

\subsubsection{DESI}

The Dark Energy Spectroscopic Instrument (DESI) is a ground-based dark energy experiment that will study baryon acoustic oscillations and the growth of structure through redshift-space distortions as a successor of the baryon oscillation spectroscopic survey (BOSS) \citep{DESI_Collaboration}.  The survey will start in 2019 mapping $1\times10^6$ of galaxies and quasars.  The full five-year DESI survey will cover a baseline survey of 14,000 deg$^2$.  DESI will be able detect quasars via Ly-$\alpha$ forest in $2.1 \lesssim z \lesssim 3.5$.  For the purpose of the Fisher analysis,  we follow the number count using DESI baseline (Fig.~3.17 in Ref.~\citep{DESI_Collaboration}) and the bias from the BOSS quasar bias $b_1(z) = 3.6 D_1(z=2.4)/D_1(z=0)$ where $D_1(z)$ is the linear growth factor \citep{Font-Ribera_ea2013}.

\begin{comment}
\begin{table}[h!]\center
\hfil\begin{tabular}{|p{0.1\tw}|p{0.15\tw}|}
	\hline
	\hfil redshift & \hfil $N(z)/\Delta z (/\text{deg}^2)$\\
	\hline
	\hfil2.0 - 2.2 & \hfil 19 \\
	\hline
	\hfil2.2 - 2.4 & \hfil 16 \\
	\hline
	\hfil2.4 - 2.6 & \hfil 12 \\
	\hline
	\hfil2.6 - 2.8 & \hfil 8 \\
	\hline
	\hfil2.8 - 3.0 & \hfil 5 \\
	\hline
\end{tabular}
\caption{DESI QSO baseline Samples with bias $b = 3.6 D(z=2.4)/D(z=0)$ where $D(z)$ is the linear growth factor.}
\end{table}
\end{comment}

\subsubsection{Our fiducial surveys}

We divide our redshift bins into three equally spaced bins as shown in Table~\ref{tab:surveys}.  In the table, the average number density of galaxies tracers $\bar{n}_g$ and the survey volume $V_s$ in Eq.~(\ref{eq:effectiveVolume}) are also shown.  The first two bins, called SKA1 and SKA2, are modeled after SKA phase 2 where we took the SKA survey parameters into account while calculating $\bar{n}_g$ and $V_s$.  Similarly, the last bin is modeled after the DESI survey.  The bias for the SKA-like and DESI-like survey are also shown.  The first two bins will provide insight to how the SKA will give constraints on a wide-area-type survey, while the last bin will represent a deep small-area-type survey.  We shall use to the mean value of the redshift in each bin,  $z = 0.5, 1.5$, and $2.5$, respectively, for the calculation of the power spectra and growth functions.

\subsection{Fisher matrix formalism}

In order to forecast the constraints on cosmological parameters for a given survey, we utilize the Fisher information matrix method \citep{Fisher_1935, Heavens2009},
\begin{equation}
	{\rm F}_{\alpha\beta} = \frac{1}{2}\mbox{Tr}\left[ \boldsymbol{\rm C}_{,\,\alpha}\, \boldsymbol{\rm C}^{-1}\, \boldsymbol{\rm C}_{,\,\beta}\, \boldsymbol{\rm C}^{-1}\right],
\end{equation}
where ${\rm F}_{\alpha\beta}$ is a component of the Fisher matrix $\boldsymbol{\rm F}$.  $\boldsymbol{\rm C}$ is the covariance matrix and ${\rm C}_{,\alpha} \equiv \partial {\rm C}/\partial \theta_\alpha$, where $\theta_\alpha$ is a parameter.  For the two-dimensional redshift-space power spectrum $P_s(k, \mu)$ a component of the Fisher matrix is given by \citep{Tegmark1997, Seo_Eisenstein2003},
\begin{eqnarray}
	\nonumber {\rm F}_{\alpha\beta} &=& \frac{1}{4\pi^2}\int_{k_{\rm min}}^{k_{\rm max}} \diff k\, k^2 \int_0^1\diff\mu \frac{\partial}{\partial \theta_\alpha}\ln P_{s,g}(k, \mu)\\[0.5em]
	&& \times\ \frac{\partial}{\partial \theta_\beta}\ln P_{s,g}(k, \mu)\, V_{\rm eff}(k, \mu).
\end{eqnarray}
The effective survey volume is given by
\begin{equation}
	\label{eq:effectiveVolume}
	V_{\rm eff}(k, \mu) = V_{\rm s} \left( \frac{\bar{n}_{\rm g} P_{\rm s, g}(k, \mu)}{1 + \bar{n}_{\rm g} P_{\rm s, g}(k, \mu)}\right)^2,
\end{equation}
where $V_{\rm s}$ is the volume of the survey and $\bar{n}_{\rm g}$ is the mean galaxy number density.  The estimated uncertainty from the inverse Fisher matrix gives an optimal uncertainties $\sigma_\alpha$ of the parameter $\theta_\alpha$ for the Cram\'{e}-Rao bound
\begin{equation}
	\sigma_\alpha^2 \ge {\rm F}^{-1}_{\alpha\alpha}.
\end{equation}
With the Fisher matrix formalism, a conservative estimate of the uncertainty in measurement could be obtained.  The matter power spectra were computed using CAMB\footnote{\url{https://camb.info/}} \citep{Lewis_ea2000}.

We shall focus our attention on the density parameters $\Omega_{\rm B} h^2, \Omega_{\rm M} h^2, \Omega_\Lambda$ and the equation of state parameters $w_0, w_{\rm a}$ as those parameters have a direct impact on the growth functions [Eqs.~(\ref{eq:growth2})-(\ref{eq:growth3-2})].  The parameters $\Delta_{\mathcal R}^2$, $\tau$, and $n_s$ are power-spectra-related parameters.  They are not directly relevant to the growth functions; however, their uncertainty in measurements will affect the constraint on other parameters.  Hence, we shall include the uncertainty in the parameters $\Delta_{\mathcal R}^2$, $\tau$, and $n_s$ in our analysis using PLANCK's priors.  The inclusion of the bias parameter $b_1$ [Eq.~(\ref{eq:bias})] is necessary as it is an unavoidable effect of galaxy clustering.  In addition, we shall assume that the bias and instrumental noise are unknown within 20\% accuracy.  It is considered as a nuisance parameter which shall be marginalized.  Hence, our fiducial cosmological parameters, after marginalization, for the Fisher analysis used in this article are
\begin{equation}
	\boldsymbol{\theta} \equiv \{\Omega_{\rm B} h^2, \Omega_{\rm M} h^2, \Omega_\Lambda, w_0, w_{\rm a}\}.
\end{equation}
The values of the fiducial cosmological parameters are summarized in Table~\ref{tab:modelsparams}.  The values of the bias parameter $b_1$ for a different model and redshift bin are summarized in Table~\ref{tab:surveys}.  In this article, all the cosmological models are assumed flat.

Our fiducial cosmological parameters for the Planck covariance matrix is based from {\tt planck\_{}lowl\_{}lowLike\_{}highL\_{}BAO} data with the base {\tt base\_{}w\_{}wa}.  This data set will give different values of $w_0$ and $w_{\rm a}$ (Eq.~(\ref{eq:planck_w0wa})).  However, we shall assume the variation of the covariance matrix is negligible across small changes in the parameter space.

\section{Results}
\label{sec:results}

In this article, the effect of dynamical dark energy on the nonlinear power spectra in the form of redshift-space distortion was investigated (see Sec.~\ref{sec:theory}).  The linear matter spectra and nonlinear matter power spectra using LPT are shown in Fig.~\ref{fig:powerspectra}.  We illustrate the observable effects of the redshift-space distortion in real space by performing the inverse Fourier transform of the nonlinear power spectra in Eq.~(\ref{eq:ps}) and the results are shown in Fig.~\ref{fig:xi}.

In order to test the model of dynamical dark energy,  we apply the Fisher matrix analysis in Sec.~\ref{sec:models} to our fiducial models.  The 68\% C.L. constraints on our parameters of interest after marginalization are shown in Table~\ref{tab:constraints}.  We divide the galaxies into different redshift bins; SKA1, SKA2, and DESI.  The details of the bins are described in Table~\ref{tab:modelsparams}.  Additionally, the combined constraints from all redshift bins (all bins) and with PLANCK's priors (all bins + PLANCK) are also shown.

\section{Discussions and Conclusions}
\label{sec:conclusion}

The nonlinear power spectra using the LPT formalism \citep{Matsubara2008} are shown in Fig.~\ref{fig:powerspectra}.  From the figure the cutoff scale, beyond which the approximation will be invalid, increases with redshift (in $k$ space).  This is due to the growth of nonlinear scales.  Even though smaller scales can be probed by high-redshift surveys, the gain of information is hindered by the decreasing the number of observed galaxies at high redshifts.  The decrement in the number of galaxies is both intrinsic and instrumental.  In Fig.~\ref{fig:bins}, the optimal bins where the constraining power is highest are the lowest bins ($z = 0.0-1.0$) and progressively decrease at higher-redshift bins (see Table~\ref{tab:surveys}).  Since all the bins are independent, we can achieve better constraints by combining the information from all the bins as shown in Fig.~\ref{fig:sigmaconstraint}.  From the figure, we could see that the constraints from the galaxy surveys alone are comparable to that of all sky PLANCK's CMB survey.  Some of the parameters have slightly better constraints than PLANCK alone.

From Fig.~\ref{fig:sigmaconstraint} and Table~\ref{tab:constraints}, the data from redshift surveys of galaxies are sensitive to $\Omega_{\Lambda}, w_0, w_{\rm a}$ and mildly sensitive to $\Omega_{\rm M} h^2$.  This is due to the fact that those parameters are directly involved in the growth functions which play an important role in late-time evolution of the Universe.  The constraint on $\Omega_{\rm B} h^2$ is not considerably better in comparison to PLANCK's constraint (see the left panel in Fig.~\ref{fig:sigmaconstraint}).  The combination of information from PLANCK with redshift surveys, in general, give tighter constraints in all the cosmological parameters.   In addition, parameter constraints from both PLANCK's cosmic microwave background observation and redshift surveys have different degeneracy directions which will give more compact constraints.  The variations of the constraining power across different models are not significantly different; however, the constraints on $w_0$ and $w_{\rm a}$ are better in the $w$CDM and $w_{\rm a}$CDM models.

Our fiducial surveys are modeled after SKA and DESI.  We apply the survey specifications from SKA for $z = 0.0 - 2.0$ and from DESI for $z = 2.0-3.0$ in Table~\ref{tab:surveys}.  SKA utilizes the 21-cm emission lines to detect HI galaxies; however, the frequency range of SKA could potentially observe the HI galaxy beyond redshift 2.0.  Since the constraining power does not significantly improve with higher redshifts, we shall only limit the SKA capability to redshift 2.0 for a conservative estimation.  The extra DESI-like deep-survey bin does not give any noticeable improvement.  We have included the uncertainties in power spectrum parameters such as $\tau$, $n_s$, and $\Delta_{\mathcal R}^2$ as the inherited uncertainty from PLANCK's CMB measurements.  We add 20\% uncertainty in the bias parameter and all the unknown instrumentation noise will be incorporated into the bias uncertainty.

\begin{figure}
	\includegraphics[width=\cw]{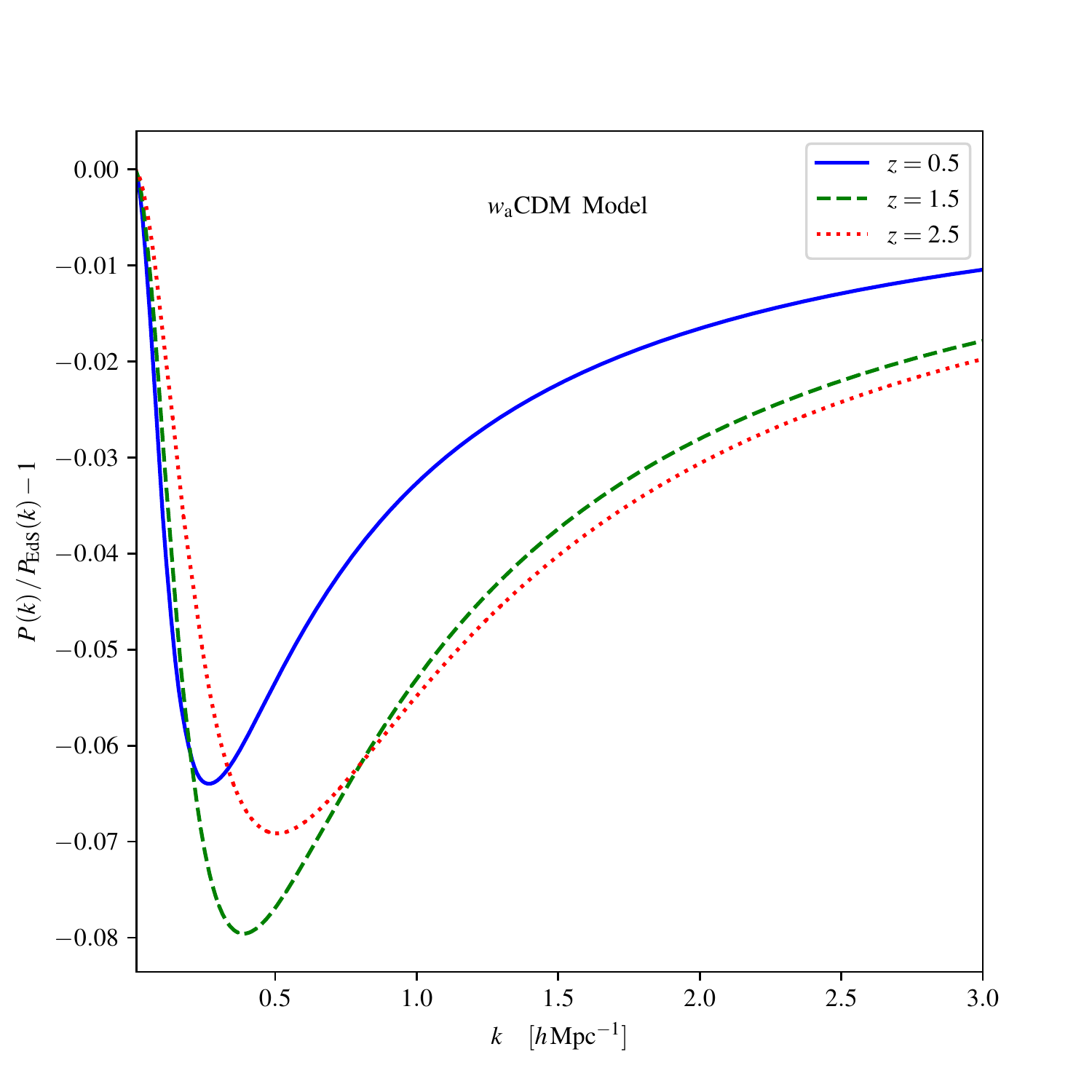}
	\caption{\label{fig:ratioPkEdS} The ratio between nonlinear power spectra with time-dependent growth functions $P(k)$ and nonlinear power spectra with an EdS approximation $P_{\rm EdS}(k)$.  The solid blue line is the nonlinear power spectra at $z = 0.5$, the dashed green line and the dotted red line are at $z = 1.5$ and $z = 2.5$ respectively.  In the figure, the model $w_{\rm a}$CDM is shown; however, the same plot for the other models has similar features.}
\end{figure}

With the instrumentation noises and the bias estimation described in the previous paragraph, the overall constraints of all the parameters (All Bins + PLANCK) are at $\sim$1\% accuracies for the density parameters, $\sim5\%$ on the equation of state parameter $w_0$ and $\sim60\%$ on $w_{\rm a}$ (see Table~\ref{tab:constraints}).  Our results are consistent with similar work done in Ref.~\citep{Bull2016}.  These precisions will rule out most of the nonstandard cosmological models and put even tighter constraints on dark energy models; therefore, an accurate theoretical model of the observables become crucial.  This is why second- and third-order growth functions [Eqs.~(\ref{eq:growth2})--(\ref{eq:growth3-2})] should be treated with care.  Many works on standard perturbation theory and Lagrangian perturbation theory use the Einstein de Sitter (EdS) values for the second- and third-order growth functions known as the \textit{EdS approximation} \citep{Jeong_Komatsu2006}.  In the EdS universe, $D_2 = (3/7)D_1^2$, $D_{3a} = (5/21)D_1^3$ and $D_{3b} = (1/6)D_1^3$ (see Sec.~\ref{sec:theory} for more details), which simply leaves only one independent growth function, namely, $D_1$.  The variations in the value of $D_2$, $D_{3a}$, and $D_{3b}$ across the realistic region in the parameter space is $\sim$1\% from that of the EdS values \citep{Bouchet_Colombi1995}.  However, the variation in the nonlinear power spectrum with time-dependent growth functions get magnified.  In Fig.~\ref{fig:ratioPkEdS}, we demonstrate the fractional difference in the nonlinear power spectra [Eq.~\ref{eq:realps}] between the EdS parameter value $P(k)$ and the non-EdS parameter value $P_{\rm EdS}(k)$.  Our result is consistent with the work in Ref.~\citep{Hiramatsu_Taruya2009}; however, our work applies the Lagrangian perturbation theory across various scales and redshifts. The non-EdS parameter values are according to the models in Table~\ref{tab:modelsparams}.  From the figure, it can be seen that the fractional difference is, on average, $\sim$5\% across interested range of redshift and could get up to $\sim10\%$ in certain redshifts.  The maximum difference occurs at $k \sim 0.5$ $h$ Mpc$^{-1}$, a quasilinear regime.  This is an indicator that as the precisions get better, the EdS approximation may not be accurate enough and a need for higher-order growth functions is emerging.

\begin{table*}
\caption{\label{tab:constraints} 68\% C.L. constraints on the fiducial cosmological parameters where $\omega_{\rm B} \equiv \Omega_{\rm B} h^2$ and $\omega_{\rm M} \equiv \Omega_{\rm M} h^2$ after marginalization.}
\begin{tabular}{|p{2em}|p{10em}||p{6em}|p{6em}|p{6em}|p{6em}|p{6em}|}
	\cline{3-7}
	\multicolumn{2}{c||}{} & \hfil $\sigma_{\omega_{\rm B}} \times 100$ & \hfil $\sigma_{\omega_{\rm M}}$ & \hfil $\sigma_{\Omega_\Lambda}$ & \hfil $\sigma_{w_0}$ & \hfil $\sigma_{w_{\rm a}}$ \\[0.5em]
	\hline
	\hline
	& \hfil SKA1 & \hfil 0.1162 & \hfil 0.00358 & \hfil 0.00248 & \hfil 0.1764 & \hfil 0.6074 \\
	\cline{2-7}
	& \hfil SKA2 & \hfil 0.1689 & \hfil 0.00571 & \hfil 0.00240 & \hfil 0.3693 & \hfil 1.2106 \\
	\cline{2-7}
	& \hfil DESI & \hfil 1.7792 & \hfil 0.05812 & \hfil 0.04672 & \hfil 5.0634 & \hfil 16.111 \\
	\cline{2-7}
	\cline{2-7}
	& \hfil All Bins & \hfil 0.0780 & \hfil 0.00233 & \hfil 0.00150 & \hfil 0.1498 & \hfil 0.5080 \\
	\cline{2-7}
	\parbox[t]{2.3em}{\multirow{-6}{*}{\rotatebox[origin=c]{90}{\small $\Lambda$CDM}}}  & \hfil All Bins + PLANCK
	& \hfil 0.0192 & \hfil 0.00058 & \hfil 0.00104 & \hfil 0.0918 & \hfil 0.3179 \\
	\hline
	\hline
	& \hfil SKA1 & \hfil 0.1180 & \hfil 0.00362 & \hfil 0.00255 & \hfil 0.2815 & \hfil 0.4355 \\
	\cline{2-7}
	& \hfil SKA2 & \hfil 0.1645 & \hfil 0.00545 & \hfil 0.00230 & \hfil 0.7768 & \hfil 1.1567 \\
	\cline{2-7}
	& \hfil DESI & \hfil 1.5174 & \hfil 0.05137 & \hfil 0.03998 & \hfil 9.4672 & \hfil 13.988 \\
	\cline{2-7}
	& \hfil All Bins & \hfil 0.0731 & \hfil 0.00220 & \hfil 0.00152 & \hfil 0.2402 & \hfil 0.3671 \\
	\cline{2-7}
	\parbox[t]{2.3em}{\multirow{-6}{*}{\rotatebox[origin=c]{90}{\small $w$CDM}}}  & \hfil All Bins + PLANCK
	& \hfil 0.0193 & \hfil 0.00064 & \hfil 0.00119 & \hfil 0.0359 & \hfil 0.0636 \\
	\hline
	\hline
	& \hfil SKA1 & \hfil 0.1259 & \hfil 0.00387 & \hfil 0.00052 & \hfil 0.3519 & \hfil 0.6101 \\
	\cline{2-7}
	& \hfil SKA2 & \hfil 0.1691 & \hfil 0.00570 & \hfil 0.00048 & \hfil 0.7394 & \hfil 1.2328 \\
	\cline{2-7}
	& \hfil DESI & \hfil 1.7552 & \hfil 0.05785 & \hfil 0.00936 & \hfil 10.640 & \hfil 17.789 \\
	\cline{2-7}
	& \hfil All Bins & \hfil 0.0825 & \hfil 0.00248 & \hfil 0.00031 & \hfil 0.2979 & \hfil 0.5121 \\
	\cline{2-7}
	\parbox[t]{2.3em}{\multirow{-6}{*}{\rotatebox[origin=c]{90}{\small $w_{\rm a}$CDM}}}  & \hfil All Bins + PLANCK
	& \hfil 0.0194 & \hfil 0.00064 & \hfil 0.00025 & \hfil 0.0361 & \hfil 0.0778 \\
	\hline
	\hline
\end{tabular}
\end{table*}

In this article, we demonstrate an application of Lagrangian perturbation formalism by Ref.~\citep{Matsubara2008} on the effects of dynamical dark energy on the observation of redshift-space power spectra and correlation functions.  With upcoming powerful telescopes such the SKA and DESI,  the constraints on the density parameters could potentially achieve better than $\sim$1\% while the constraints on dark energy parameters $w_0$ and $w_{\rm a}$ could be better than $\sim5\%$ and $\sim60\%$.  However, a better, more accurate theoretical modeling is also needed.  The EdS approximation which is often used for cosmological perturbation theory may need to be revised.  We suggest that the inclusion of the time-dependent growth functions in the second-order and third-order term become crucial as the precision get below $1\%$; however, the impact of these functions on the outcome of measurements still needs further investigation.

\begin{acknowledgments}
	We would like to thank to Sirichai Chongchitnan, Chris Gordon, and Pimpunyawat Tummuangpak for their helpful suggestions, and Utane Sawangwit for his useful help in the initial part of the project and the High Performance Computer facility at National Astronomical Research Institute of Thailand (NARIT) where some parts of the calculation were done.   T.C. acknowledges support from the Thailand Research Fund (TRF) through Grant No.~MRG550041.
\end{acknowledgments}

%\bibliography{lpt_refs}

\end{document}